\documentclass{PoS}
\newcommand{\les}{\stackrel{<}{{}_{\sim}}}

\title{
\vspace{-1.5cm}
\hspace{13.0cm}{\small\rm KEK-CP-336}\\[5pt]
Study of the conformal region of the SU(3) gauge theory with domain-wall fermions}

\ShortTitle{Study of the conformal region of the SU(3) gauge theory with domain-wall fermions}

\author{\speaker{Junichi Noaki}\\%
        KEK High Energy Accelerator Research Organization \\
        E-mail: \email{noaki@post.kek.jp}}

\author{Guido Cossu\\
        KEK High Energy Accelerator Research Organization \\
        E-mail: \email{cossu@post.kek.jp}}

\author{Ken-Ichi Ishikawa\\
     Hiroshima University, Department of Physical Science \\
      E-mail: \email{ishikawa@theo.phys.sci.hiroshima-u.ac.jp}}

\author{Yoichi Iwasaki \\
       University of Tsukuba \\
       E-mail: \email{iwasaki@ccs.tsukuba.ac.jp}}

\author{ Tomoteru Yoshi\'e \\
       University of Tsukuba \\
       E-mail: \email{yoshie@ccs.tsukuba.ac.jp}}

\abstract{ We investigate the phase structure of the SU(3) gauge theory with  $N_f=8$ 
 by numerical simulations employing the massless Domain-Wall fermions.
 Our aim is to study directly the massless quark region, since it is the most important  region to
 clarify the properties of conformal theories.
 When the number of flavor is within the conformal window, it is claimed recently with Wilson quarks that there is 
the conformal region at the small quark mass region in the parameter space in addition to the confining phase and the deconfining phase.
 We study the properties of the conformal region 
investing the spatial Polyakov loops and the temporal meson propagators.
Our data imply that there is the conformal region, and a phase transition between the confining phase
 and the conformal region takes place. These results are consistent with the claim that the conformal window is between $7$ and $16$.
Progress reports on other related studies are also presented.}

\FullConference{The 33rd International Symposium on Lattice Field Theory\\
		 14 -18 July  2015\\
		 Kobe International Conference Center, Kobe, Japan}

\begin{document}

\section{Introduction}

\begin{figure}[t]
 \begin{center}
  \includegraphics[width=8.0cm,clip]{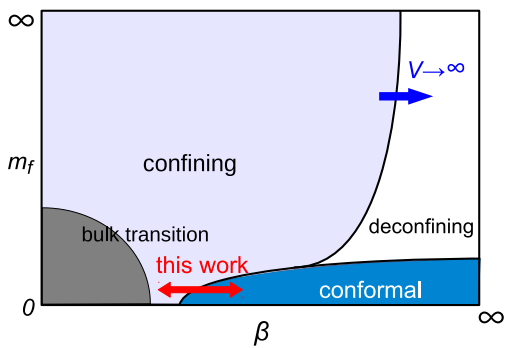} 
 \end{center}
 \vspace{-1.1cm}
 \caption{Sketch of the phase structure in the ($\beta,m_f$)-plane.}
\end{figure}

The $SU(3)$ gauge theories with $N_f$-fermions in the fundamental
representation which we call ``many-flavor conformal QCD'' are believed to have 
conformal properties when $N_f$ is within the conformal window. This model have 
been intensively studied 
in the context of the composite Higgs model as a candidate of the beyond
standard model (See for a review \cite{HasenfrazReview}.)
Besides the phenomenological interest, it is theoretically interesting and important to clarify the properties of the theory. 
In refs.~\cite{ConformalIR,GlobalStr}  the authors claim, based on the study with the Wilson fermions, that
\begin{enumerate}
 \item the conformal window is $7\leq N_f\leq 16$

 \item  in addition to the confining phase and the deconfining phase there is a conformal region for 
 $m_f \leq \Lambda_{\rm IR}$ (the IR-cutoff) in the deconfining phase

  \item in the conformal region the vacuum is the non-trivial twisted $Z(3)$ vacuum modified non-perturbatively

 \item in the conformal region meson propagators exhibit power corrected Yukawa type decay
 \end{enumerate}

Our objective is in this article to investigate these claims by numerical simulations with the massless $N_f=8$ 
Domain-Wall fermions.
For this purpose, the phase diagram predicted in ref.~\cite{GlobalStr} 
is translated to 
the one shown in Figure~1, where $m_f$ is the fermion mass.
In the light fermion region of the diagram, the conformal phase 
is separated from the confining phase at a certain value of $\beta$.
On the other hand, in the heavier mass region, the deconfining phase 
would disappear in the limit of infinite lattice volume.
The use of massless Domain-Wall fermions allows us to study the light 
mass region without tuning the fermion mass to zero as in the Wilson formalism.
The fact that the chiral symmetry is realized to a good approximation
is also advantageous to address the question whether or not the
conformal theory is chiral symmetric.

After presenting the details of our numerical simulations in Section 2,
we show, in Section 3, the vacuum structure based on the calculation 
of the spatial Polyakov loops and the meson temporal propagators.
Section 4 contains progress report on the study 
of identifying the location of the IR fixed point using a scaling properties of meson propagators.
We summarize our future plan in Section 5.

\section{Numerical Simulation}

\begin{table}[t]
\begin{center}
  \begin{minipage}{0.42\textwidth}
   $V=8^3\times 32$\\
   \begin{tabular}{lllll}
    \hline
    \, $\beta$&\, $\Delta\tau$&\, \#traj.&\, AC&\ \ $am_{\rm res}$  \\
    \hline\hline
     2.0 &0.25&\, 8,000&\ \, 500& $2.2\times 10^{-4}$\\
     2.3 &0.25& 12,000 & 1,000& $2.7\times 10^{-5}$\\
     2.6 &0.10& 11,000 & 2,500& $1.2\times 10^{-5}$\\
     4.2 &0.25& 10,000 & 1,500& $1.3\times 10^{-6}$\\
     4.7 &0.25&\, 9,000& 1,500& $1.2\times 10^{-6}$\\
     6.0 &0.25&\, 9,000& 1,500& $8.9\times 10^{-7}$\\
    \hline
   \end{tabular}
  \end{minipage}
\hspace{1.0cm}
  \begin{minipage}{0.42\textwidth}
   $V=16^3\times 64$\\
   \begin{tabular}{lllll}
    \hline
    \, $\beta$&\, $\Delta\tau$&\, \#traj.&\, AC&\ \ $am_{\rm res}$  \\
    \hline\hline
     2.6&0.100&\, 5,500 &\ \, 500 & $1.0\times 10^{-5}$\\
     3.4&0.125&\, 8,000 &1,000& $2.8\times 10^{-6}$\\
     4.2&0.125&12,000 &1,200& $1.5\times 10^{-6}$\\
     4.7&0.125&11,000 &1,500& $1.3\times 10^{-6}$\\
     6.0&0.125&11,000 &1,500& $1.0\times 10^{-6}$\\
    \hline
   \end{tabular}
   \vspace*{0.45cm}
  \end{minipage}
\end{center}
 \caption{Profiles of the generated gauge ensembles. 
 For the lattice volume $V=8^3\times 32$ (left) and $16^3\times 64$
 (right), values of $\beta$, MD-step size $\Delta\tau$, numbers of 
 thermalized trajectories, 
 a rough estimation of the auto-correlation (AC) time
 and typical values of the residual fermion mass are listed for each.}
\end{table}

In our numerical simulation, we employ the Iwasaki action for the gauge
sector and the $N_f=8$ Domain-Wall action for the fermion.
The Domain-Wall formalism we use is the same as the one originally proposed 
by Shamir in order to avoid complication related to its improvements.
Throughout this study, the bare fermion mass is set to zero.
The Domain-Wall height and the 5-th dimensional width are also fixed
to $M=1.6$ and $L_s=12$, respectively.
We set periodic boundary conditions in the spatial directions and
an anti-periodic condition in the temporal direction.
By using the IroIro++ code system~\cite{IroIroProc}, we generate gauge 
configurations with several values of $\beta$ on
two different volumes $8^3\times 32$ and $16^3\times 64$. 
The unconventionally large size in the time
direction is for a critical study of the meson propagation presented 
in Section~4.
In the molecular dynamics of HMC, we use the two time-step Omelyan 
integrator with the step size $\Delta\tau$ and $\Delta\tau/4$ 
for pseudo-fermions and link variables, respectively.
In Table~1, $\Delta\tau$ and the number of thermalized trajectories
are listed for each ensemble ($\beta$ and the lattice volume).
The acceptance rates are more than 90\% for all generations.

Because of the massless simulation, slow thermalization
and long auto-correlations are anticipated, especially for
the large $\beta$ ensembles.
We estimate these MD-times by monitoring the magnitude of spatial 
Polyakov loops $|P_x|$, $|P_y|$ and $|P_z|$ on the configurations. 
From a perturbative argument of the vacuum structure~\cite{GlobalStr},
at the larger 
$\beta$ region, the spatial Polyakov will converge to a non-zero value (with some fluctuation) whereas 
the temporal ones remain around zero.
We can exploit the condition that  the spatial Polyakov loops converge to some non-zero values. 
Results from these rough estimation are listed in the third 
and fourth column of the tables. It is noted that the number of thermalized 
trajectories are several times larger than the auto-correlation length 
for all gauge ensembles. Therefore, the statistical localization can be 
avoided by measurements covering whole thermalized trajectories.

To check the chiral property of the generated configurations, 
we compute residual fermion mass $am_{\rm res}$ 
using the definition by the violation of the Ginsparg-Wilson 
relation~\cite{DWF}.
Simply averaging the measured values on a few sample configurations for 
each ensemble, we obtain typical magnitudes of $am_{\rm res}$ as
listed in the last column.
Those values are sufficiently small ({\it i.e.} 0.1\% of the pseudo-scalar 
masses on the corresponding ensembles) to allow us to directly investigate
the properties of the conformal region in the diagram.

\section{Vacuum structure}

\begin{figure}[t]
 \begin{center}
  \includegraphics[width=4.8cm]{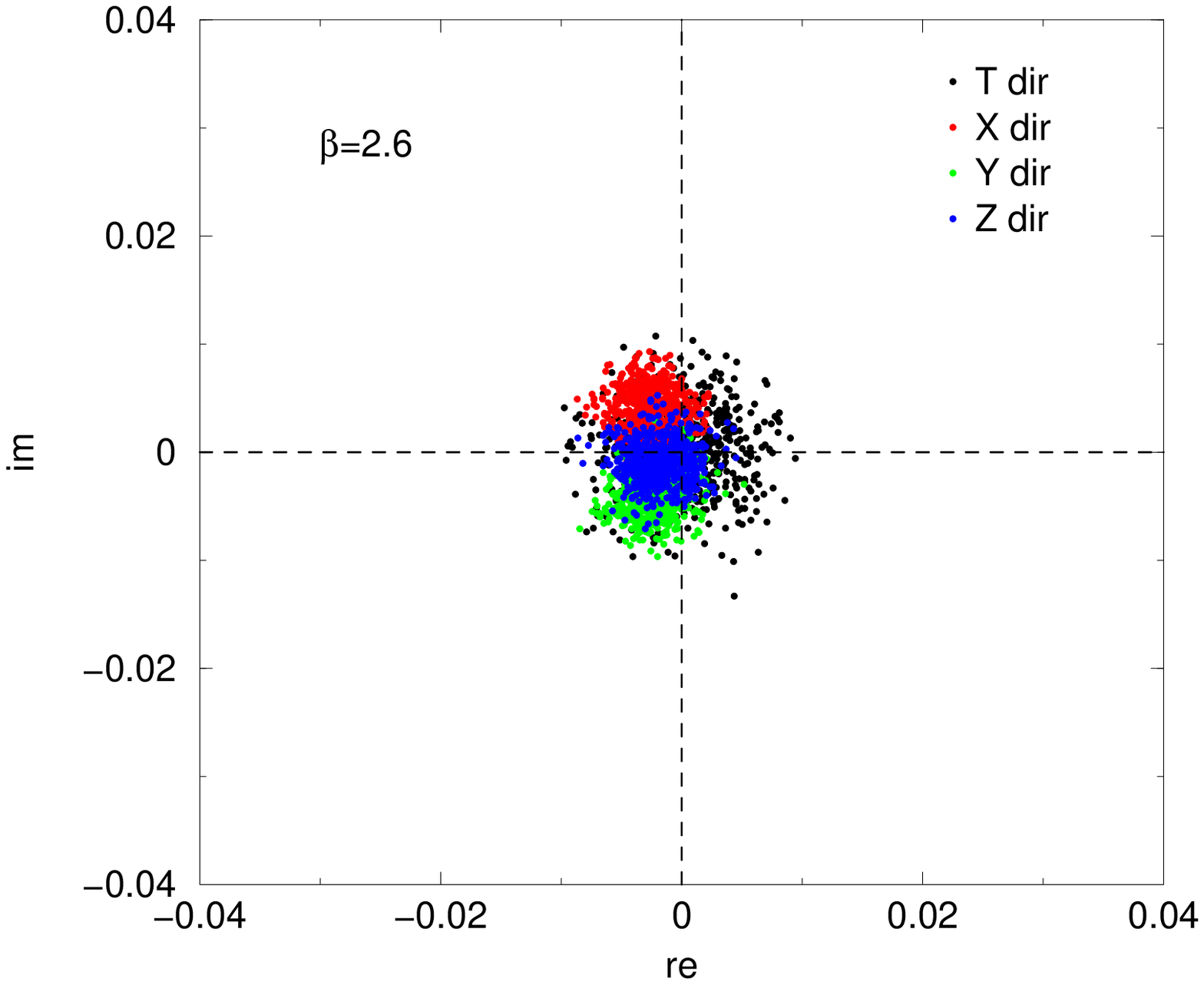}
  \includegraphics[width=4.8cm]{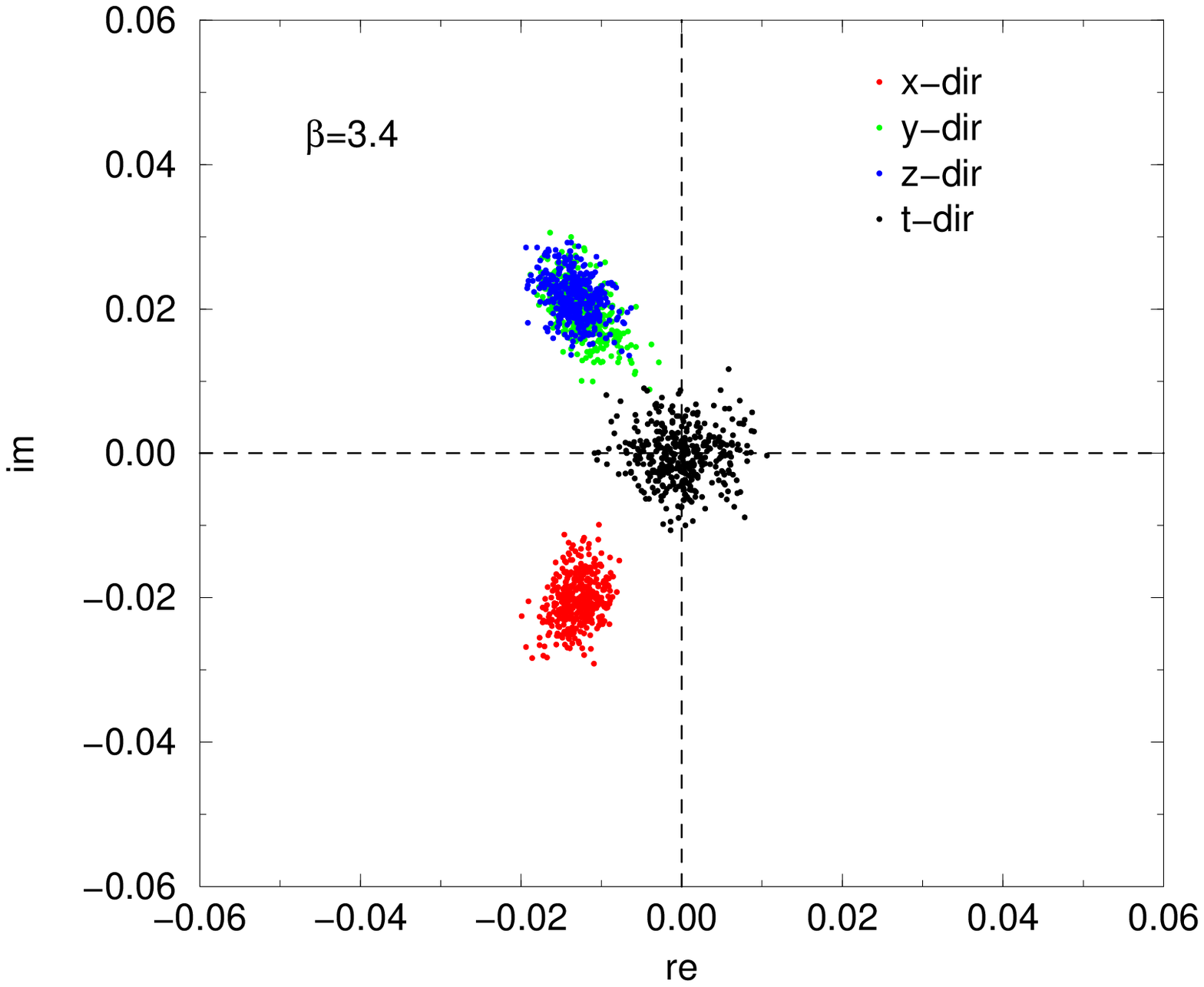}
  \includegraphics[width=4.8cm]{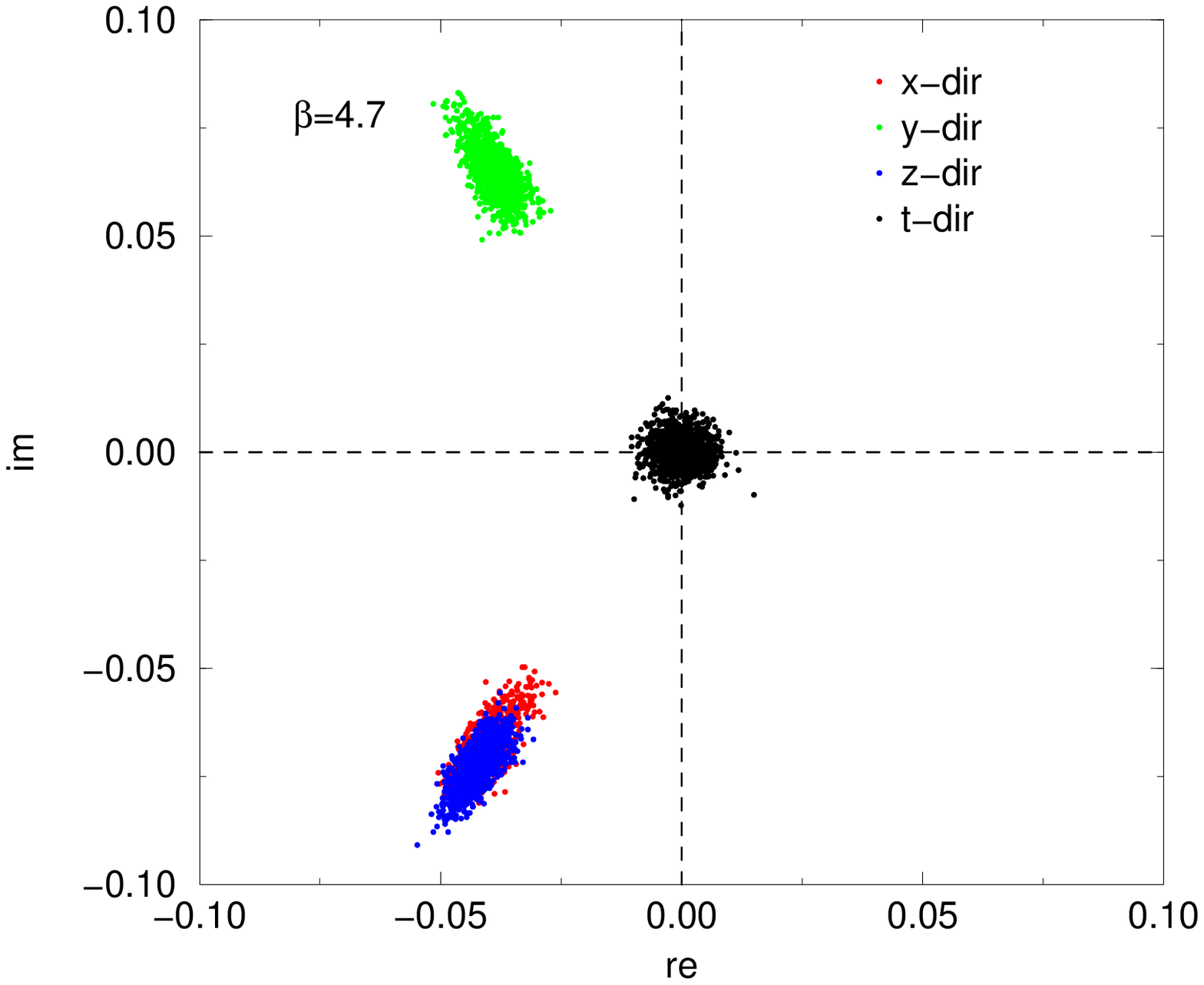}
 \end{center}
 \caption{Distribution of the Polyakov loop measured on each gauge
 configurations at $\beta=2.6$ (left),$3.4$ (center) and $4.2$ (right).
 In each panel, $P_t$ is shown by black symbols and $P_{x,y,z}$ by 
 different colors. Results on the $16^3\times 64$ lattice are shown.}
\end{figure}

\begin{figure}[t]
 \begin{center}
  \includegraphics[width=4.9cm]{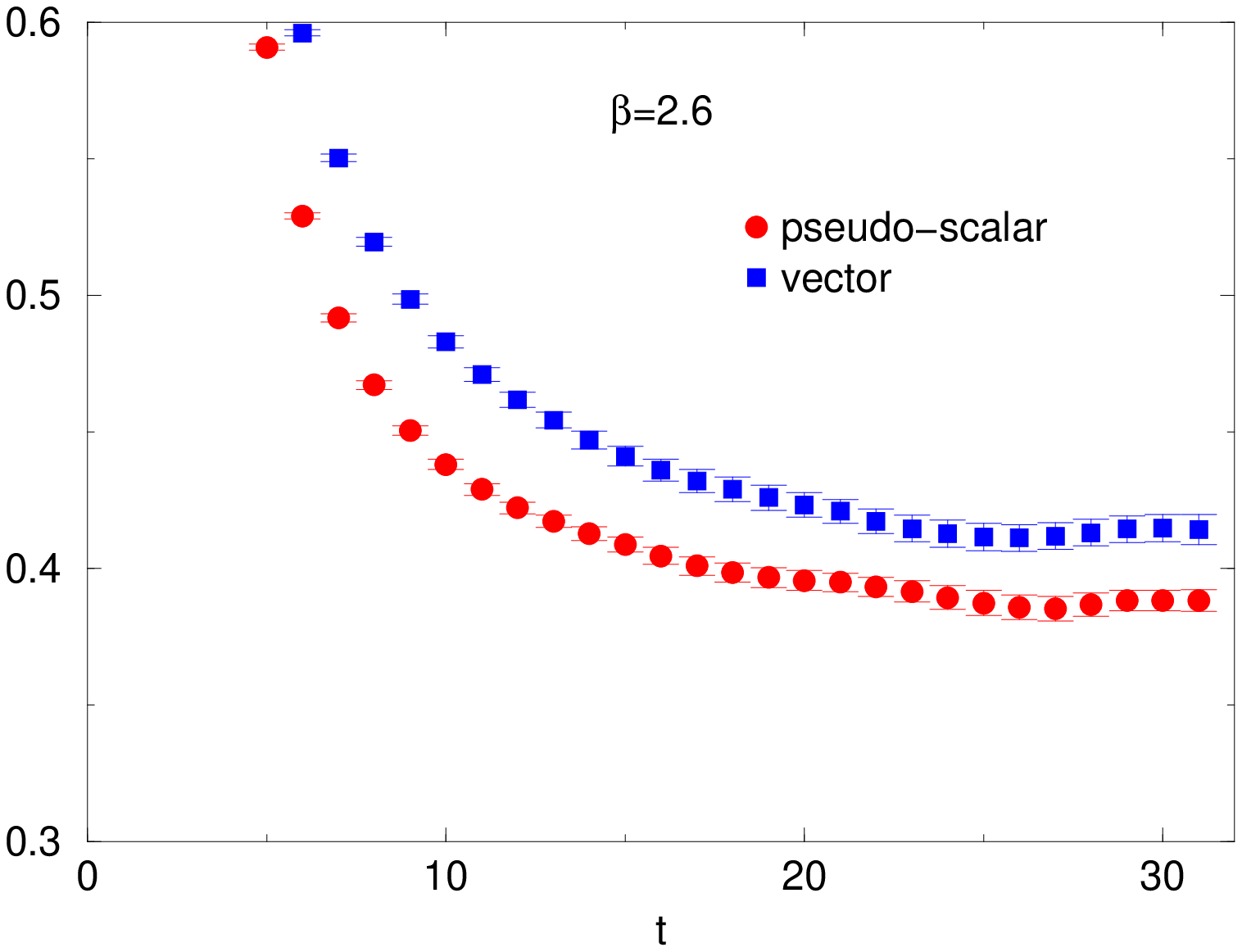}
  \includegraphics[width=4.9cm]{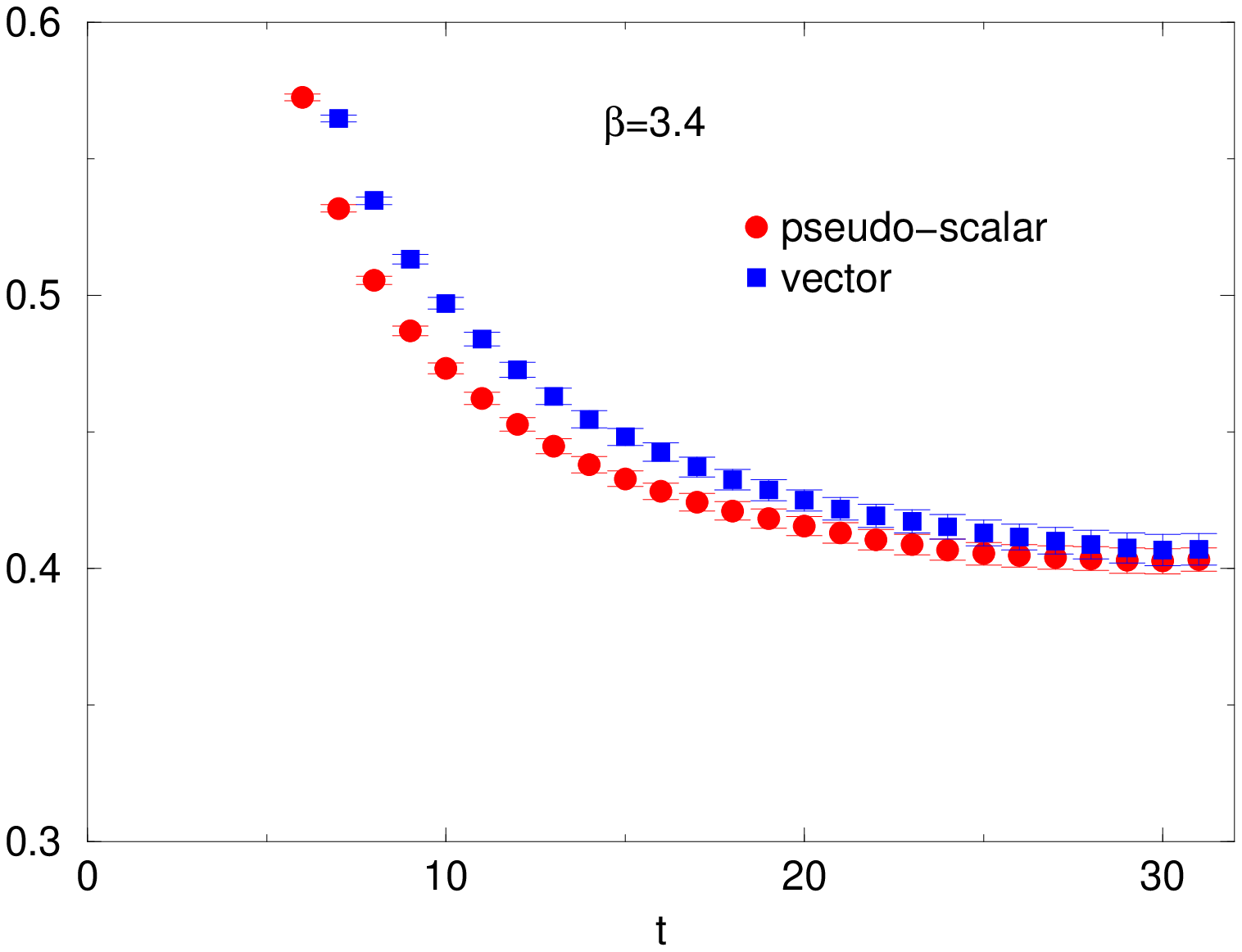}
  \includegraphics[width=4.9cm]{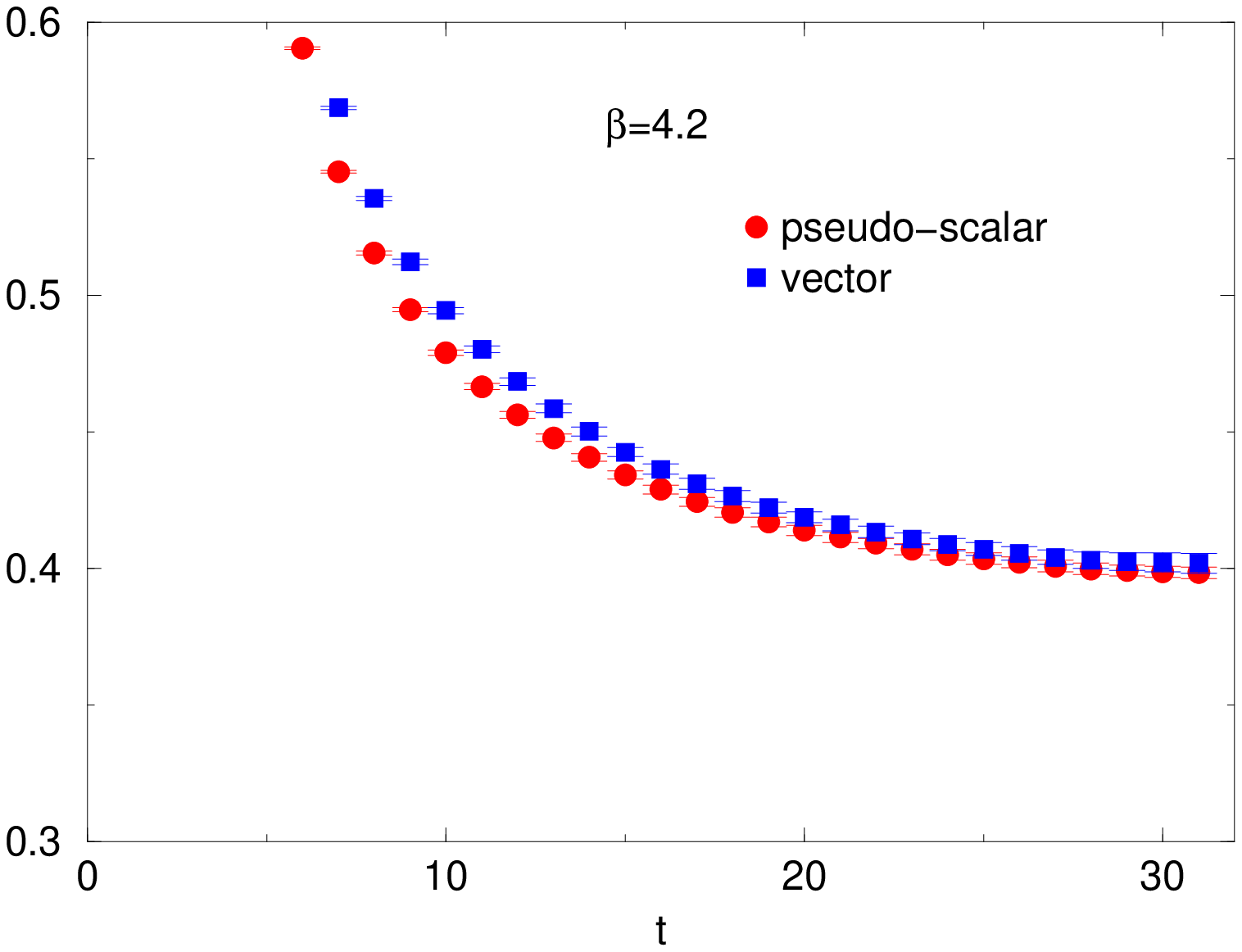}
 \end{center}
 \caption{Effective masses of pseudo-scalar (red) and vector (blue)
 mesons at $\beta=2.6$ (left),$3.4$ (center) and $4.2$ (right).
Results on the $16^3\times 64$ lattice are shown.}
\end{figure}

Figure 2 shows the distribution of Polyakov loops in the complex plane
at different $\beta$'s on the $16^3\times 64$ lattice.
We observe that at $\beta=2.6$ the distribution is centralized at the origin, while at $\beta=3.4$ 
the Polyakov loops in the spatial directions take the angles $\pm 2\pi/3$, as in the previous works with the Wilson fermions~\cite{GlobalStr}. At $\beta=4.2,$ the distribution becomes sharper and the magnitude increases.
Thus the $Z(3)$ symmetry is broken for $\beta \geq 3.4$ in the spatial
directions, but the $Z(3)$ symmetry is preserved in the temporal direction. 
In other words, 
a phase transition from the $Z(3)$ 
symmetry symmetry phase to the broken phase seems to occur
between $\beta=2.6$ and $\beta=3.4$.

The meson temporal propagators and masses contain useful and important
information 
for the investigation of the properties of the vacuum structure.
The correlation functions are computed in steps of 10 trajectories. 
We improve the signal by taking average of the propagator obtained with 
local sources at different time-slices in intervals of 
8 for $\beta=2.6$ and $3.4$ on the $16^3\times 64$ lattice, 
and of 4 for others. 
Three panels of Figure~3 show the effective masses obtained 
assuming the $\cosh$ function on the $16^3\times 64$ lattice. 
In each panel, we show the vector and pseudo-scalar channels at the same
$\beta$'s as Figure~2. The mass difference seen at $\beta=2.6$
diminishes at $\beta=3.4$ and disappears at even higher $\beta$.
This is a sign of restoration of the chiral symmetry.
Combined with the analysis of the spatial Polyakov loops above
the data suggest that 
 there is a phase transition taking place between the confining phase
 and the conformal phase between $\beta=2.6$ and $\beta=3.4$.

We note that the data behavior described in this section is more 
distinct with the $16^3\times 64$ volume than $8^3\times 32$.
This implies that data obtained in a even larger volume would 
allow us to investigate properties of the transition in more detail.
In the following discussion, we assume our gauge ensembles 
at $\beta \ge 3.4$ on the $16^3\times 64$ are in the conformal region.

\section{Meson propagators}

\begin{figure}[t]
 \begin{center}
  \includegraphics[width=6.8cm]{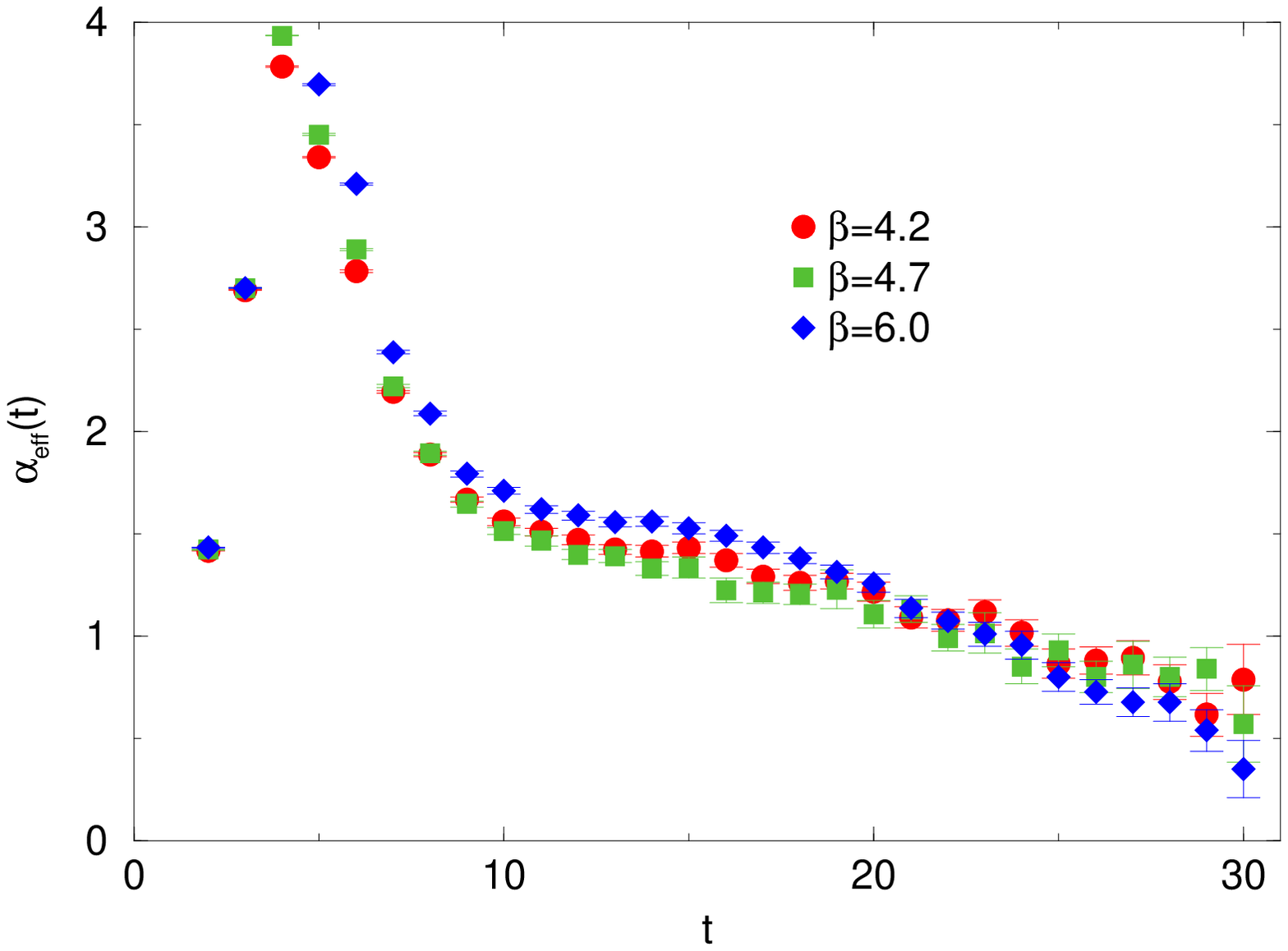}
  \includegraphics[width=6.8cm]{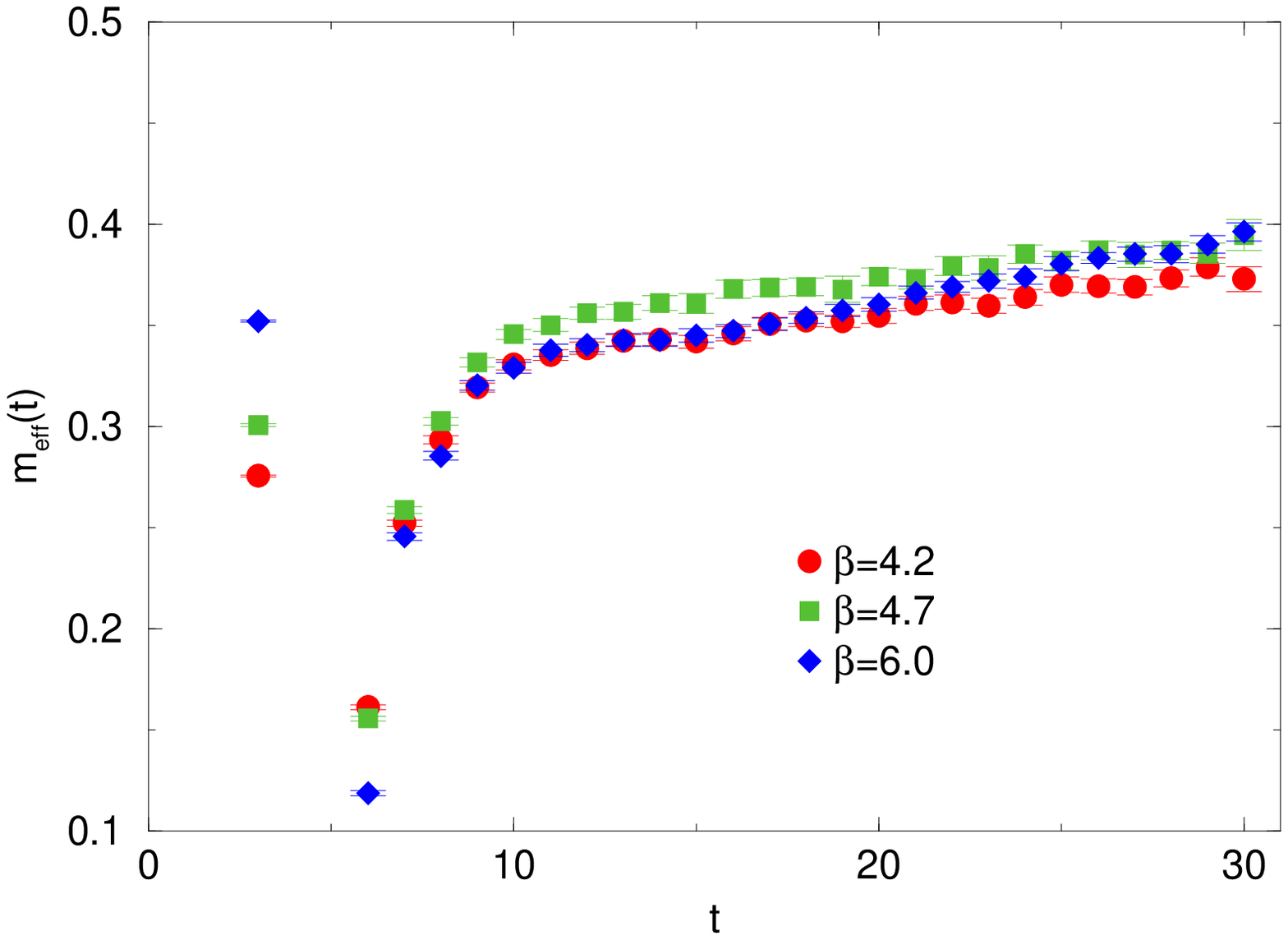}
 \end{center}
 \vspace*{0.4cm}

 \caption{Effective values of $\alpha$ (left panel) and $m$ 
 (right panel) on the $16^3\times 64$ lattice.
 Results at $\beta=4.2$ (red), $4.7$ (green) and $6.0$ (blue)
 are shown.}
\end{figure}

Now we investigate the property of the conformal phase through 
the pseudo-scalar meson temporal propagator.
It is known that in the conformal theory all correlation functions without any IR cutoff
show a power law behavior, $\propto t^{-\alpha}$,
at long distances. However, given our system is in a finite box, 
the propagation is influenced by the IR-cutoff. Ref.~\cite{GlobalStr} 
predicts that the modified behavior is 
$\propto t^{-\alpha} e^{-mt}$ (``the power-law corrected Yukawa type
decay''), which looks a hybrid of the power law and the ordinary
exponential decay, 
$\propto e^{-mt}.$

We parametrize the propagator as $G(t) = At^{-\alpha} e^{-mt}$.
Assuming a mild $t$-dependence of $\alpha$ and $m$ such that 
they do not significantly change between time slices at $t$ and $t\pm 1$,
those effective values are obtained from the data of $G(t)$ and $G(t\pm 1)$.
In the two panels 
of Figure~4, we plot those quantities on the $16^3\times 64$ lattice for 
$\beta=4.2,\ 4.7$ and $6.0$ with different colors.
In the figure, the assumption of the mild $t$-dependence is
reasonably satisfied while there is no significant $\beta$-dependence
observed with the current statistics. 
Applying the relation $\alpha=3 - 2 \gamma*_m$ to the plat region $10 \leq t \leq 20$ for $\beta=4.2$, 
we obtain a rough estimation 
$0.2 \les \gamma_m \les 0.6$, which is consistent with the previous result~\cite{GlobalStr}
 (see Eq. $11$, pages $31$ and $33$). To increase the reliability of the estimate we will increase statistics at $\beta=3.4$,
 which is closer to the transition point. (We have excluded the data at $\beta=3.4$ in Figure~4 due to poor statistics.)

\begin{figure}[t]
 \begin{center}
  \includegraphics[width=5.1cm]{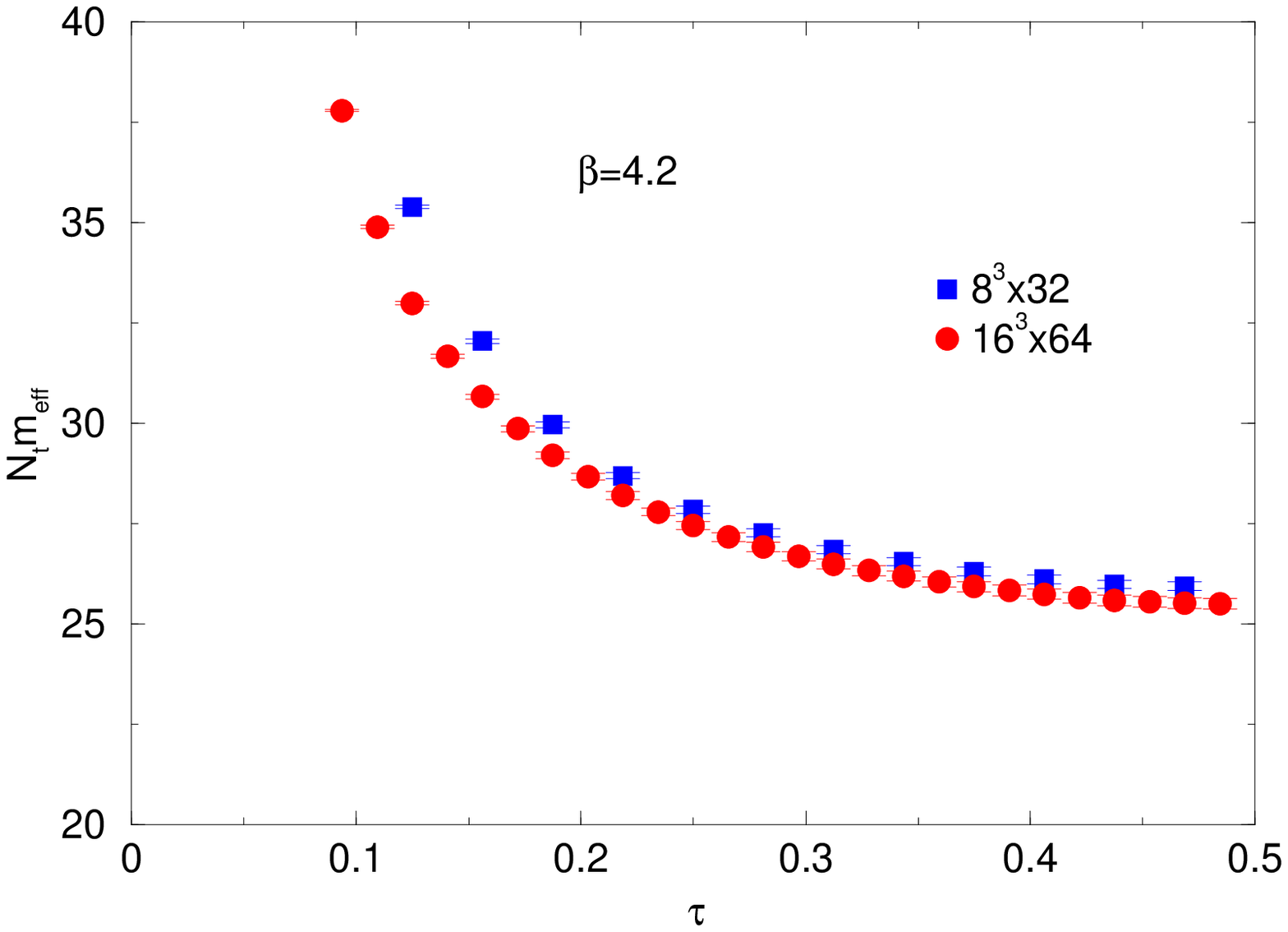} 
  \hspace{0.1cm}
  \includegraphics[width=4.8cm]{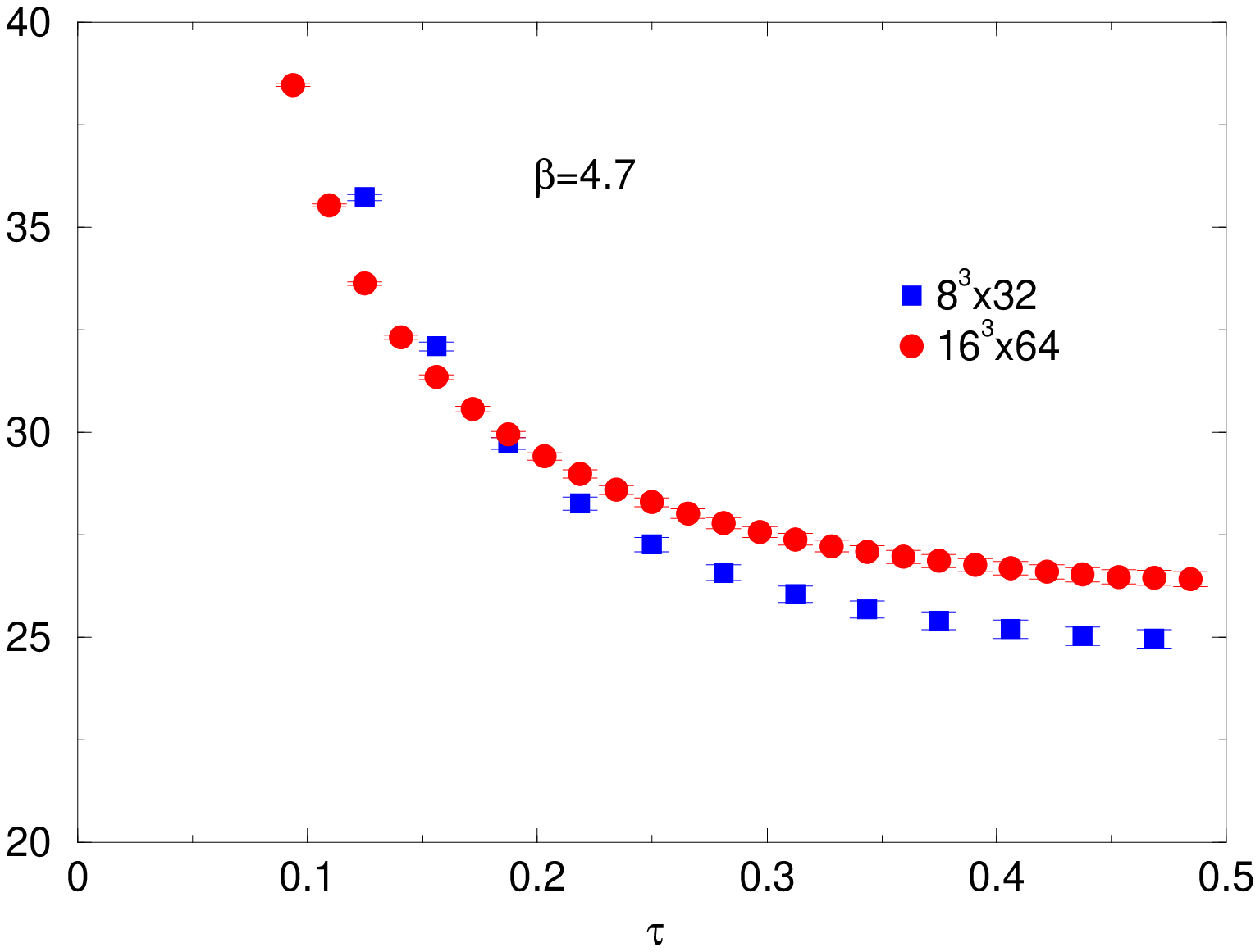} 
  \includegraphics[width=4.8cm]{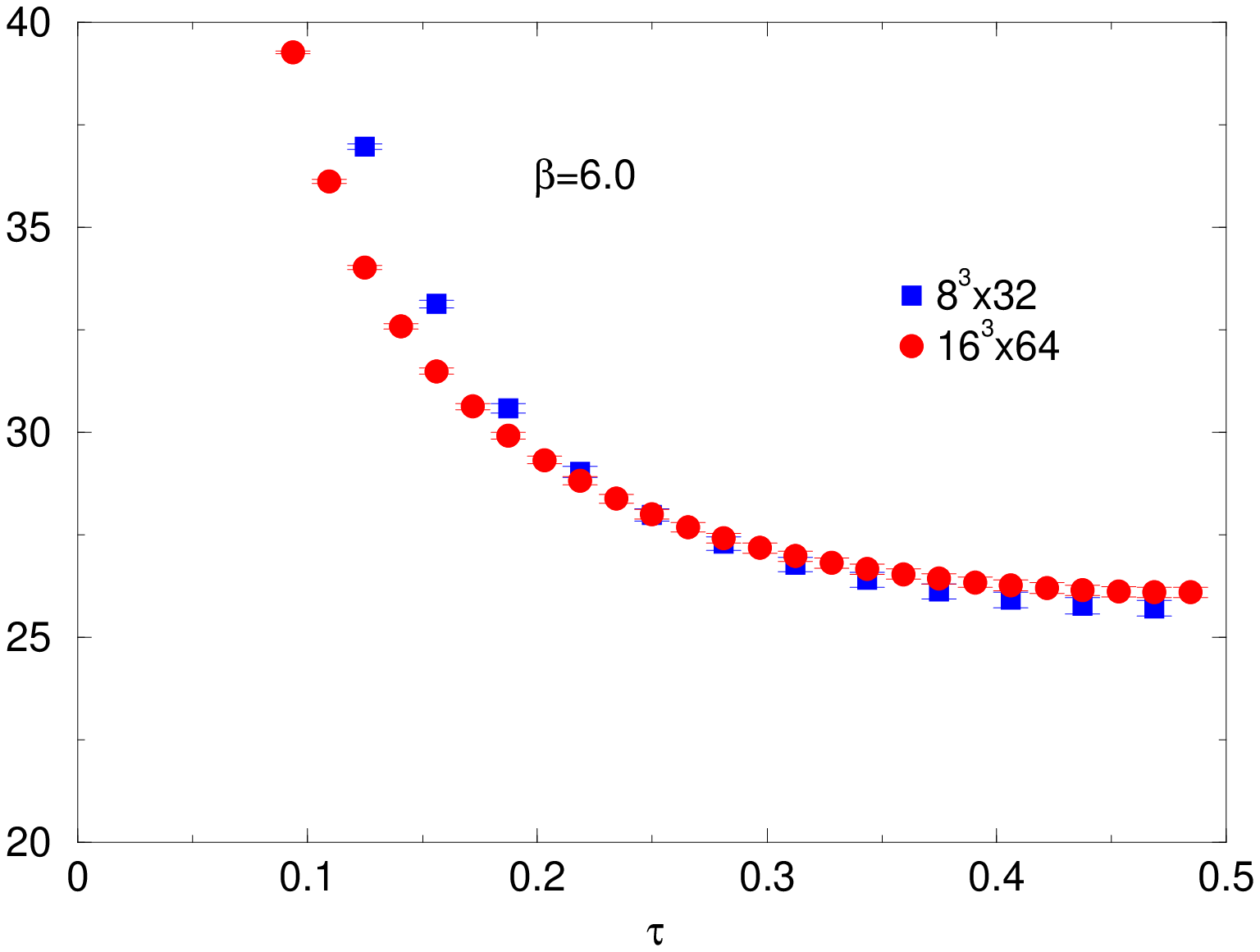} 
 \end{center}
 \vspace{0.2cm}
 \caption{ Comparison of the effective mass $\hat{m}$ obtained 
 with two different volumes $16^3\times 64$ (red) and $8^3\times 32$
 (blue)  at $\beta= 4.2$ (left), $4.7$ (center) and $6.0$ (right).}
\end{figure}

As another important analysis using the meson propagator, we attempt
to determine the location of the infra-red fixed point (IRFP)~\cite{IRfixedPt}.
For the lattice size $N$, we consider the scaled propagator 
$\tilde{G}(\tau=t/N_t ; N) =G(t ; N)$.
At IRFP, due to the finite size scaling, $\tilde{G}$ with 
the different lattice sizes $N$ and $N'$ are related by
\begin{eqnarray}
 \tilde{G}(\tau ; N) = (N'/N)^{3-2\gamma_m}\tilde{G}(\tau ; N').
\end{eqnarray}
This relation further implies that the scaled effective mass defined by
\begin{eqnarray}
\hat{m}(\tau; N)\equiv N_t\ln(G(t; N)/G(t+1; N))
\end{eqnarray}
 should be volume-independent at IRFP. 
In Figure~5, we compare $\hat{m}$ at three different $\beta$'s in the conformal region on the two lattice volumes.
In the figure, we observe the magnitude relation of the two volumes at
large $\tau$ changes between $\beta=4.2$ and $4.7$, which implies the existence 
of IRFP in this region. 
For a conclusive determination of the existence and the
location of IRFP, we need more data points with different $\beta$ 
to narrow the $\beta$-region and those with larger volume to confirm the 
size-independence.
 
\section{Future plans}

Besides the analyses presented in the last two sections, we attempt
to obtain spectral density $\rho(\lambda)$ of the Dirac-eigenvalue
$\lambda$ by the stochastic method~\cite{GiustiLuscher,Holland,Hashimoto}.
However, the discussion based on the Banks-Casher relation requires 
the large volume limit before the chiral limit is taken.
Also, it should be noted that the sign of the chiral symmetry violation 
may be so delicate that a careful study as seen in ref.~\cite{CossuEtAl} 
is necessary.
That means calculations have to be done in a sufficiently large volume
with non-zero fermion masses to be able to determine whether 
or not the chiral symmetry is conserved in the conformal region. 
The same is valid to extract $\gamma_m$ in a reliable manner.

As seen in the plots presented in Section 3 and 4, 
the data of meson propagator have to be improved by more statistics.
We are also planning to generate gauge ensembles on a larger volume
to conclude our study. 
Combining the data on those ensembles with the current ones, 
it is possible to sort out the issues of the phase transition in Section 3 
and of the property of the meson propagator in Section 4,
as well as the above Dirac-spectrum analysis.
A natural target is $32^3\times 128$. 
At the same time, we aim to understand the relation of our work with
others which use different fermion formalism or setup~\cite{Boulder,KMI,KMI2}.

\section*{Acknowledgment}
 Numerical simulations are performed on the IBM System Blue Gene Solution 
 at High Energy Accelerator Research Organization (KEK) under a support of 
 its Large Scale Simulation Program (No. 14/15-18). 
 This work is supported in part by the Grant-in-Aid of the Japanese 
 Ministry of Education (No. 15K05065) and the SPIRE 
 (Strategic Program for Innovative Research) Field5 project.

\newcommand{\J}[4]{{#1} {\bf #2} (#3) #4}
\newcommand{\RMP}{Rev.~Mod.~Phys.}
\newcommand{\MPL}{Mod.~Phys.~Lett.}
\newcommand{\IJMP}{Int.~J.~Mod.~Phys.}
\newcommand{\NP}{Nucl.~Phys.}
\newcommand{\NPSup}{Nucl.~Phys.~{\bf B} (Proc.~Suppl.)}
\newcommand{\PL}{Phys.~Lett.}
\newcommand{\PRD}{Phys.~Rev.~D}
\newcommand{\PRL}{Phys.~Rev.~Lett.}
\newcommand{\AP}{Ann.~Phys.}
\newcommand{\CMP}{Commun.~Math.~Phys.}
\newcommand{\CPC}{Comp.~Phys.~Comm.}
\newcommand{\PTP}{Prog. Theor. Phys.}
\newcommand{\Suppl}{Prog. Theor. Phys. Suppl.}
\newcommand{\JHEP}{JHEP}
\newcommand{\PoS}{PoS}

\end{document}